\documentclass[aps,prb,twocolumn,amsmath,amssymb,nofootinbib,
  superscriptaddress]{revtex4-2}

\usepackage{graphicx}
\usepackage{bm}
\usepackage{amsmath}
\usepackage{amssymb}
\usepackage{physics}
\usepackage[colorlinks=true,linkcolor=blue,urlcolor=blue,citecolor=blue]{hyperref}
\usepackage[usenames,svgnames]{xcolor}
\usepackage{makecell}
\usepackage[normalem]{ulem}

\usepackage{simplewick}  
\usepackage{simpler-wick}
\usepackage{booktabs}

\newcommand{\Hop}{\hat{H}}

\newcommand{\Himp}{\hat{H}_{\rm imp}}
\newcommand{\gHimp}{\mathcal{H}_{\rm imp}}

\newcommand{\Hbath}{\hat{H}_{\rm bath}}
\newcommand{\Hhyb}{\hat{H}_{\rm hyb}}

\newcommand{\aop}{\hat{a}}
\newcommand{\adop}{\hat{a}^{\dagger}}

\newcommand{\nbar}{\bar{n}}

\newcommand{\cop}{\hat{c}}
\newcommand{\cdop}{\hat{c}^{\dagger}}
\newcommand{\hc}{{\rm H.c.}}
\newcommand{\rhoop}{\hat{\rho}}
\newcommand{\Zimp}{Z_{{\rm imp}}}

\newcommand{\gK}{\mathcal{K}}
\newcommand{\gI}{\mathcal{I}}
\newcommand{\gF}{\mathcal{F}}
\newcommand{\bolda}{\bm{a}}
\newcommand{\boldabar}{\bar{\bm{a}}}
\newcommand{\abar}{\bar{a}}
\newcommand{\im}{{\rm i}}

\newcommand{\gA}{\mathcal{A}}

\newcommand{\WI}{W^I}
\newcommand{\WII}{W^{II}}

\newcommand{\ustc}{State Key Laboratory of Precision and Intelligent Chemistry, University of Science and Technology of China, Hefei 230026, China}

\newcommand{\nudt}{College of Science, National University of Defense Technology, Changsha 410073, China}

\begin{document}


\title{Improved time-translationally invariant tensor network influence functional method for Anderson impurity problems}

\author{Zhijie Sun}
\affiliation{\ustc}


\author{Zhenyu Li}
\email{zyli@ustc.edu.cn}
\affiliation{\ustc}

\author{Chu Guo}
\email{guochu604b@gmail.com}

\affiliation{\nudt}



\begin{abstract}
The Anderson impurity model (AIM) is of fundamental importance in condensed matter physics for studying strongly correlated phenomena. However, accurately simulating its long-time dynamics still remains a significant numerical challenge. A class of recently developed numerical approaches represents the Feynman-Vernon influence functional (IF), which encodes all the bath effects on the impurity, as a matrix product state (MPS) in the temporal domain. The computational cost of this approach is largely determined by the bond dimension $\chi$ of the temporal MPS.
In this work, we propose an efficient and accurate method that, when the hybridization function in the IF can be approximated as a sum of $n$ exponential functions, 
systematically constructs the IF as an MPS by multiplying $O(n)$ small MPSs, each with bond dimension $2$. 
Our method yields a worst case scaling of $\chi$ as $2^{8n}$ and $2^{2n}$ for real- and imaginary-time evolution respectively. 
We demonstrate the performance of our method for two commonly used bath spectral functions, and show that the required bond dimensions are significantly smaller than the worst case. 
\end{abstract}

\maketitle


\section{Introduction}

The Anderson impurity model describes a localized impurity coupled to a continuous, noninteracting bath~\cite{anderson1961-localized}, and serves as a prototypical model for studying a wide range of quantum phenomena, including strongly correlated effects~\cite{mahan2000-many,Georges2016}, quantum transport~\cite{MeisnerDagotto2009,SchwarzWeichselbaum2018}, and open quantum dynamics~\cite{weiss1993-quantum}. The AIM and its multi-orbital generalizations also constitute the central problem in dynamical mean field theory (DMFT) and its non-equilibrium extensions for high-dimensional quantum many-body systems~\cite{MetznerVollhardt1989,GeorgesKotliar1992,GeorgesRozenberg1996,KotliarMarianetti2006}.
Owing to its fundamental importance, a wide variety of numerical methods have been developed to solve the AIM, including exact diagonalization~\cite{CaffarelKrauth1994,KochGunnarsson2008,GranathStrand2012,LuHaverkort2014,ZaeraLin2020,HeLu2014,HeLu2015}, continuous-time quantum Monte Carlo~\cite{RubtsovLichtenstein2004,RubtsovLichtenstein2005,WernerMillis2006,GullTroyer2008,ChanMillis2009,haule2010-dynamical,GullWerner2011,huang2014-electronic,lu2016-pressure,yue2021-pairing}, numerical renormalization group~\cite{Wilson1975,Bulla1999,BullaPruschke2008,Frithjof2008,ZitkoPruschke2009,DengGeorges2013,StadlerWeichselbaum2015,LeeWeichselbaum2016,LeeWeichselbaum2017}, time-evolving MPS~\cite{WolfSchollwock2014b,GanahlEvertz2014,GanahlVerstraete2015,WolfSchollwock2015,GarciaRozenberg2004,NishimotoJeckelmann2006,WeichselbaumDelft2009,BauernfeindEvertz2017,LuHaverkort2019,WernerArrigoni2023,KohnSantoro2021,KohnSantoro2022,WautersBurrello2023,WautersBurrello2024}, hierarchical equation of motion (HEOM)~\cite{YoshitakaKubo1989,jin2007-dynamics,jin2008-exact,yan2016-dissipation,cao2023-recent}, and the pseudomode theory~\cite{TamascelliPlenio2018,ThoennissAbanin2024,ParkLin2024,HuangLin2026}. 
However, despite their success in certain regimes or for certain types of impurity problems, accurately and efficiently simulating real- and long-time dynamics remains a significant challenge.

The AIM possesses a particularly favorable mathematical structure: the impurity is linearly coupled to a non-interacting bath. When the interaction within the impurity vanishes, the entire model becomes quadratic and thus exactly solvable. Even in the interacting case, the model still admits a partial analytical treatment via the Feynman-Vernon IF, in which the bath degrees of freedom are integrated out, leaving only the impurity degrees of freedom defined along the time contour~\cite{FeynmanVernon1963}. 

An emerging numerical approach for studying the AIM is to represent the IF as an MPS, thereby fully exploiting this partial analytic structure. This approach offers several advantages over conventional methods: (1) it is free from bath discretization errors, as the bath degrees of freedom are integrated out analytically; (2) it does not suffer from the sign problem, as is typical for MPS-based algorithms; and (3) it is likely to be computationally more efficient, as only the impurity degrees of freedom along the time contour are retained in the IF. Within this framework, two distinct approaches have been developed to date, which we collectively refer to as tensor network IF method. The first and pioneering approach represents the IF as an MPS in the Fock state basis and has been applied to the AIM on both the real-time~\cite{ThoennissAbanin2023a,ThoennissAbanin2023b,NgReichman2023,ParkChan2024} and imaginary-time~\cite{KlossAbanin2023} axis. The second approach represents the IF as a Grassmann MPS (GMPS) in the coherent state basis. This formulation can be viewed as a direct fermionic analogue of the time-evolving matrix product operator (TEMPO) method~\cite{StrathearnLovett2018}, and will hereafter be referred to as the Grassmann TEMPO (GTEMPO) method. GTEMPO has been applied to the AIM and their multi-orbital extensions on the real-time~\cite{ChenGuo2024a,ChenGuo2024c}, imaginary-time~\cite{ChenGuo2024b} and the L-shaped Kadanoff contours~\cite{ChenGuo2024g}. A more detailed comparison between these two approaches is presented in the next section.

In this work, we propose an efficient and accurate method to construct the IF as a GMPS, which is a central step in the GTEMPO method.
The starting point of our method is to approximate the hybridization function in the IF as a sum of $n$ exponential functions, in close analogy to HEOM and the pseudomode theory.
The contribution of each exponential term can be represented as a GMPS with bond dimension $2$, using the standard $\WII$ algorithm~\cite{ZaletelPollmann2015}, which is a first-order approximation in general.
The IF can then be constructed as a GMPS by multiplying $2n$ such small GMPSs for imaginary-time evolution, or $8n$ for real-time evolution. 
Crucially, we prove that in this setting the $\WII$ approximation becomes exact, owing to the special structure of Grassmann algebra.
As a result, our method achieves a lower computational cost than existing alternatives that have been used in GTEMPO.
Our method yields a worst-case scaling of the bond dimension $\chi$ as $2^{2n}$ and $2^{8n}$ for imaginary- and real-time evolution, respectively. For real-time evolution, it has been shown that, under mild assumptions on the hybridization function, the number of exponentials scales as $n\sim\log(t)\log(1/\varepsilon)$ for a total evolution time $t$ and a prescribed error tolerance $\varepsilon$~\cite{VilkoviskiyAbanin2024,ThoennissAbanin2024,HuangLin2026b}. Consequently, the overall computational cost of our method grows only polynomially with $t$, even in the worst case.
We benchmark the performance of our method using two commonly employed bath spectral functions (BSFs), namely the semi-circular and Lorentzian function, and compare the results with those obtained from the two existing approaches used in GTEMPO.

This paper is organized as follows. 
In Sec.~\ref{sec:background}, we briefly review the background relevant to our method.
Sec.~\ref{sec:method} presents the details of the proposed approach, along with its connections to and distinctions from TEMPO and the alternative tensor network IF methods. In Sec.~\ref{sec:result}, we present numerical results for two commonly used BSFs to demonstrate the efficiency and accuracy of our method. Finally, we conclude in Sec.~\ref{sec:summary}.

\section{Background}\label{sec:background}
\subsection{The path integral formalism}
The total Hamiltonian of the AIM can be written as
\begin{align}
\Hop = \Himp + \Hbath + \Hhyb,
\end{align}
where the impurity Hamiltonian $\Himp = \epsilon_d \sum_{\sigma=\uparrow\downarrow} \adop_{\sigma}\aop_{\sigma} + U\adop_{\uparrow}\adop_{\downarrow}\aop_{\downarrow}\aop_{\uparrow}$ describes a localized electron orbital with on-site energy $\epsilon_d$ and interaction strength $U$ (with $\adop$, $\aop$ denoting fermionic creation and annihilation operators, respectively). The bath Hamiltonian $\Hbath = \sum_{k, \sigma}\epsilon_k \cdop_{k, \sigma}\cop_{k, \sigma}$ describes non-interacting itinerant electrons with dispersion $\epsilon_{k}$, The hybridization term $\Hhyb = \sum_{k, \sigma}\nu_k(\adop_{\sigma}\cop_{k, \sigma} + \hc)$ accounts for the coupling between the impurity and the bath, with hybridization strength $\nu_k$. The BSF $J(\epsilon)$ is defined as
\begin{align}\label{eq:Jw}
J(\epsilon) = \sum_k \nu_k^2 \delta(\epsilon - \epsilon_k),
\end{align}
which fully characterizes the influence of $\Hhyb$ and $\Hbath$ on the impurity dynamics.

The linear structure of $\Hhyb$ (more precisely, its linearity in the bath creation and annihilation operators), together with the non-interacting form of $\Hbath$, ensures that the bath degrees of freedom can be analytically integrated out in the total path integral, leading to a reduced path integral for the impurity partition function $\Zimp$ that involving only the impurity degrees of freedom along the time axis.
For imaginary- and real-time evolution, the impurity partition functions are defined as $\Zimp(\beta) = \Tr e^{-\beta\Hop}/\Tr\rhoop^{\rm th}_{\rm bath}$ and $\Zimp(t) = \Tr\rhoop(t)/\Tr\rhoop^{\rm th}_{\rm bath}$, respectively. Here $\beta$ denotes the inverse temperature, $\rhoop^{\rm th}_{\rm bath} = e^{-\beta\Hbath}$ is the thermal state of the bath, and $\rhoop(t)$ is the total density matrix at time $t$.
In the following, we focus on the imaginary-time path integral formalism and briefly comment on the corresponding real-time formulation when necessary. For more detailed derivations and explicit expressions on different contours, we refer the reader to Refs.~\cite{XuChen2024,Chen2025}.

The impurity partition function of the AIM for imaginary-time evolution can be written as:
\begin{align}\label{eq:PI}
    \Zimp(\beta) = \int \mathcal{D}[\boldabar \bolda] \gK[\boldabar \bolda] \prod_{\sigma} \gI_{\sigma}[\boldabar_{\sigma} \bolda_{\sigma}],
\end{align}
where $\boldabar_{\sigma} = \{\abar_{\sigma}(\tau)\}$, $\bolda_{\sigma} = \{a_{\sigma}(\tau)\}$ denote Grassmann trajectories. For brevity, we have use $\boldabar = \{\boldabar_{\uparrow}, \boldabar_{\downarrow}\}$ and $\bolda = \{\bolda_{\uparrow}, \bolda_{\downarrow}\}$. The integration measure is defined as
\begin{align}\label{eq:measure}
\mathcal{D}[\boldabar \bolda] = \prod_{\sigma, \tau}\dd \abar_{\sigma}(\tau)\dd a_{\sigma}(\tau)e^{-\abar_{\sigma}(\tau)a_{\sigma}(\tau)}.
\end{align}
$\gK$ describes the contribution from the bare impurity dynamics governed by $\Himp$, and can be formally written as
\begin{align}\label{eq:gK}
\gK[\boldabar \bolda] = e^{- \int_0^{\beta} \dd\tau \gHimp(\tau)}, 
\end{align}
where $\gHimp(\tau)$ is obtained from $\Himp$ by making the substitutions $\aop_{\sigma}\rightarrow a_{\sigma}(\tau)$, $\adop_{\sigma}\rightarrow \abar_{\sigma}(\tau)$. The influence of $\Hhyb$ and $\Hbath$ is encoded in the IF $\gI_{\sigma}$, which can be written as
\begin{align}\label{eq:gI}
\gI_{\sigma}[\boldabar_{\sigma}\bolda_{\sigma}] = e^{-\int_0^{\beta}\dd \tau'\int_0^{\beta} \dd \tau'' \abar_{\sigma}(\tau')\Delta(\tau', \tau'')a_{\sigma}(\tau'')}.
\end{align}
Here $\Delta(\tau', \tau'')$ is the hybridization function, defined as
\begin{align}\label{eq:Delta}
\Delta(\tau', \tau'') = \int \dd\epsilon J(\epsilon)D_{\epsilon}(\tau', \tau''),
\end{align}
where $D_{\epsilon}(\tau', \tau'')$ is the free bath Matsubara Green’s function, defined as
\begin{align}\label{eq:Dtau}
D_{\epsilon}(\tau', \tau'') = -[\Theta(\tau' - \tau'') - \nbar(\epsilon)]e^{-\epsilon(\tau' - \tau'')},
\end{align}
with $\Theta$ the Heaviside step function and $\nbar(\epsilon) = (e^{\beta\epsilon}+1)^{-1}$ the Fermi–Dirac distribution. The integrand of the impurity path integral in Eq.(\ref{eq:PI}) is often referred to as the augmented density tensor (ADT), denoted as
\begin{align}\label{eq:adt}
\gA[\boldabar \bolda] = \gK[\boldabar \bolda] \prod_{\sigma} \gI_{\sigma}[\boldabar_{\sigma} \bolda_{\sigma}].
\end{align}

The corresponding expressions for the real-time impurity partition function $\Zimp(t)$ can be obtained by replacing the imaginary-time contour in Eq.(\ref{eq:gK}) and Eq.(\ref{eq:gI}) with the Keldysh contour~\cite{Keldysh1965}, and by substituting the free bath Matsubara Green’s function in Eq.(\ref{eq:Dtau}) with the corresponding contour-ordered Green’s function.
We further note that, for both imaginary- and real-time evolution, the IF is time-translationally invariant (TTI), i.e., $\Delta(\tau',\tau'') = \Delta(\tau'-\tau'')$, which follows from the time-independence of $\Hhyb$ and $\Hbath$. $\gK$ is also TTI provided that $\Himp$ is time-independent.

\subsection{Brief review of the GTEMPO method}

The seminal work that directly translate the analytic expression of the IF into a numerical algorithm is the quasi-adiabatic propagator path integral (QuAPI) method~\cite{makarov1994-path,makri1995-numerical}. QuAPI was originally developed for bosonic impurity problems, but its central idea can be straightforwardly generalized to the fermionic case~\cite{Chen2025}. In the fermionic case and for imaginary-time evolution, one first discretizes the Grassmann trajectories into a set of discrete Grassmann variables (GVs): $\boldabar_{\sigma} \rightarrow  \{ \abar_{\sigma,M}, \cdots, \abar_{\sigma,1} \}$ and $\bolda_{\sigma} \rightarrow  \{ a_{\sigma,M}, \cdots, a_{\sigma,1} \}$. For notational simplicity, we continue to use $\boldabar_{\sigma}$ and $\bolda_{\sigma}$ to denote these collections of discrete GVs in the following. The time grid is chosen to be uniform, with step size $\delta\tau = \beta/(M-1)$. 
A first-order approximation to $\gI_{\sigma}$ then reads
\begin{align}\label{eq:IF2}
\gI_{\sigma}[\boldabar_{\sigma}\bolda_{\sigma}] \approx e^{-\sum_{j,k=1}^M \abar_{\sigma, j}\Delta_{j, k}a_{\sigma, k}},
\end{align}
where 
\begin{align}
\Delta_{j, k} = \int_{j\delta\tau}^{(j+1)\delta\tau}\dd\tau'\int_{k\delta\tau}^{(k+1)\delta\tau}\dd\tau'' \Delta(\tau', \tau'').
\end{align}
The bare impurity contribution $\gK$ can be discretized in a similar manner, yielding a numerically exact expression up to first-order Trotter errors. Since the present work focuses on the treatment of the IF, which constitutes the computationally dominant part of the AIM, we do not elaborate on the discretization of $\gK$ here and refer the reader to Ref.~\cite{ChenGuo2024b} for details.
In our implementation, we adopt the ordering of the discrete GVs as $a_M \bar{a}_M \dots a_2 \bar{a}_2 a_1 \bar{a}_1$ when constructing each $\gI_{\sigma}$ and $\gK$ as GMPSs.

We note that the discretized $\gI_{\sigma}$ in Eq.(\ref{eq:IF2}) is a Grassmann tensor (GT) defined on $2M$ GVs. Since all terms in the exponent of the discretized $\gI_{\sigma}$ commute with each other, a straightforward construction of this GT proceeds as follows: each factor $e^{-\abar_{\sigma, j}\Delta_{j, k}a_{\sigma, k}} = 1 - \abar_{\sigma, j}\Delta_{j, k}a_{\sigma, k}$ can be represented as a two-variate GT, and the full tensor is then obtained by multiply $M^2$ such GTs together~\cite{XuChen2024} (the GT multiplication corresponds to the element-wise multiplication of the normal tensors in the bosonic case~\cite{GuoChen2024d}). In a similar manner, $\gK$ can also be constructed as a GT of $2M$ GVs. The ADT is then constructed by multiplying $\gK$ and $\gI_{\sigma}$ according to Eq.(\ref{eq:adt}). Once the ADT is obtained, (multi-time) impurity observables can be evaluated straightforwardly. For instance, the impurity partition function $\Zimp(\beta)$ is obtained by integrating out all the GVs with the measure in Eq.(\ref{eq:measure}), and the Matsubara Green's function can be calculated as
\begin{align}\label{eq:Matsubara}
G(j\delta\tau) &= \langle \aop_{\sigma, j+1}\adop_{\sigma, 1}\rangle \nonumber \\ 
&= \Zimp^{-1}(\beta)\int \mathcal{D}[\boldabar \bolda] a_{\sigma, j+1}\abar_{\sigma, 1} \gA[\boldabar \bolda],
\end{align}
where $\langle\cdots\rangle$ in the first line denotes the thermal average.

An obvious obstacle of the above approach is that the number of GVs grows linearly with the number of time steps $M$, leading to an exponential growth in the size of the ADT. To overcome this difficulty, a memory truncation scheme is introduced in QuAPI for real-time evolution, in which only the ADT within a finite time window $[j\delta t, (j+M_c)\delta t]$ is retained, where the memory size $M_c$ fixes the size of the ADT. However, such a hard memory cutoff can lead to several issues. For instance, it may introduce uncontrolled error, and it is not applicable to imaginary-time evolution, since $\Delta(\tau', \tau'')$ does not necessarily decay with the time separation $|\tau'' - \tau'|$~\cite{GuoChen2024f}.

The TEMPO method represents a significant advancement over QuAPI~\cite{StrathearnLovett2018}. Rather than representing the ADT as a single tensor, it utilizes an MPS representation, reformulating necessary tensor operations (such as element-wise multiplication) into corresponding MPS operations. A key advantage of TEMPO is its capability to accommodate substantially larger values of $M_c$, This is because the computational cost of MPS algorithms scales only linearly with the number of sites (albeit cubically with the bond dimension). Consequently, TEMPO does not strictly require an explicit memory cutoff, although the original work retained this approximation to further enhance computational efficiency.
The GTEMPO method is a fermionic analog of TEMPO, in which the discretized $\gK$ and $\gI_{\sigma}$ are represented as GMPSs rather than standard MPSs~\cite{ChenGuo2024a}. Crucially, the GMPS structure naturally and automatically account for the sign factors arising from the exchange of GVs.

In the following, we provide a more detailed discussion of the differences and connections between GTEMPO and TEMPO, as well as between GTEMPO and alternative tensor network IF methods for fermionic impurity problems. 
The original QuAPI and TEMPO typically consider a specific type of hybridization Hamiltonian, namely $\Hhyb =\hat{X}\sum_{k}\nu_k(\hat{b}_k^{\dagger}+\hat{b}_k)$ where $\hat{b}_k^{\dagger}$ and $\hat{b}_k$ are bosonic creation and annihilation operators. This form is commonly referred to as \textit{diagonal coupling}. since it involves only a single Hermitian impurity operator $\hat{X}$, one can evaluate the impurity path integral in the eigenbasis of $\hat{X}$. This yields a bosonic IF similar in form to Eq.(\ref{eq:gI}), namely,
\begin{align}\label{eq:IFb}
\gI[\bold{s}] = e^{-\int_0^{\beta}\dd \tau'\int_0^{\beta} \dd \tau'' s(\tau')\Delta(\tau', \tau'')s(\tau'')},
\end{align}
where $s(\tau)$ denotes an eigenvalue of $\hat{X}$. However, this approach becomes inapplicable when multiple non-commutative impurity operators are coupled to the same bath — a scenario known as \textit{off-diagonal coupling}. A simple example of such off-diagonal coupling is $\Hhyb' =\sum_{k}\nu_k(\hat{X}\hat{b}_k^{\dagger}+\hat{X}^{\dagger}\hat{b}_k)$ with $\hat{X}^{\dagger} \neq \hat{X}$.
Solutions for such $\Hhyb'$ have only recently been developed in Refs.~\cite{Link2026,GuoChen2026a}. For instance, the central idea in Ref.~\cite{GuoChen2026a} is to formulate a generalized Feynman-Vernon IF using operator paths $\hat{X}^{\dagger}(\tau)$ and $\hat{X}(\tau)$, rather than scalar variable path $s(\tau)$:
\begin{align}\label{eq:IFb2}
\gI[\hat{\bold{X}}^{\dagger}\hat{\bold{X}}] = e^{-\int_0^{\beta}\dd \tau'\int_0^{\beta} \dd \tau'' \hat{X}^{\dagger}(\tau')\Delta(\tau', \tau'')\hat{X}(\tau'')}.
\end{align}
After discretization, $\gI[\hat{\bold{X}}^{\dagger}\hat{\bold{X}}]$ takes the form of a matrix product operator (MPO) rather than an MPS.

In the AIM (or its multi-orbital extensions), the hybridization Hamiltonian is inherently off-diagonal, as the two impurity operators $\aop_{\sigma}$ and $\adop_{\sigma}$ do not commute. 
Drawing a parallel to Eq.(\ref{eq:IFb2}), one may attempt to convert the Grassmann expression in Eq.(\ref{eq:gI}) into an expression of normal fermionic operators via the mapping $\abar_{\sigma}\rightarrow \adop_{\sigma}$ and $a_{\sigma}\rightarrow \aop_{\sigma}$, such that one can use standard fermionic tensor network algorithms (see Ref.~\cite{BultinckVerstraete2017} for example) to build the resulting IF as a fermionic MPO in the Fock state basis. Unfortunately, such a formal generalization is incorrect in general, as the commutation relations of the GVs are different from those of the corresponding fermionic operators. For example, we have $\{\adop_{\sigma}(\tau), \aop_{\sigma}(\tau)\} = \adop_{\sigma}(\tau) \aop_{\sigma}(\tau) + \aop_{\sigma}(\tau)\adop_{\sigma}(\tau)=1$ and $[\adop_{\sigma}(\tau)\aop_{\sigma}(\tau)]^2 = \adop_{\sigma}(\tau)\aop_{\sigma}(\tau)$, while $\{\abar_{\sigma}(\tau), a_{\sigma}(\tau)\}=0$ and $[\abar_{\sigma}(\tau)a_{\sigma}(\tau)]^2 = 0$.

A pioneering approach for representing the IF of the AIM as a temporal MPS was proposed in Ref.~\cite{ThoennissAbanin2023b}. In this method, one performs a particle-hole transformation $\abar_{\downarrow, i} \rightarrow a_{\downarrow, M+i}$ in $\gI_{\downarrow}$ and $a_{\uparrow, i}\rightarrow \abar_{\uparrow, M+i}$ in $\gI_{\uparrow}$ within Eq.(\ref{eq:IF2}). This ensures that the transformed $\tilde{\gI}_{\downarrow}$ contains only $\bolda_{\downarrow}$, while $\tilde{\gI}_{\uparrow}$ contains only $\boldabar_{\uparrow}$. Consequently, $\tilde{\gI}_{\sigma}$ can be exactly mapped into a fermionic Gaussian state:
\begin{align}\label{eq:abaninIF}
e^{-\sum_{j,k=1}^M \abar_{\uparrow, j}\Delta_{j, k}\abar_{\uparrow, M+k}} \rightarrow  e^{-\sum_{j,k=1}^M \adop_{\uparrow, j}\Delta_{j, k}\adop_{\uparrow, M+k}}\vert \emptyset\rangle.
\end{align}
since the anti-commutation relations within the exponents are identical on both sides ($\vert \emptyset\rangle$ denotes the fermionic vacuum state). 
This Gaussian state has a Bardeen-Cooper-Schrieffer form, and one could use existing fermionic tensor network techniques to build it as a fermionic MPS.
The central idea of this approach was subsequently adopted by Ref.~\cite{NgReichman2023}, which employed a different derivation without using the particle-hole transformation but ultimately arrived at an expression similar to Eq.(\ref{eq:abaninIF}).

The GTEMPO method takes a more direct approach: because the IF in Eq.(\ref{eq:gI}) already provides an analytic expression in the coherent state basis in terms of Grassmann trajectories, one can directly construct it as a GMPS without mapping the GVs to fermionic operators. During the construction of the GMPS and the ADT, essentially only a single operation is required, namely, the multiplication of two GMPSs, an operation that naturally stems from the multiplication of two GTs~\cite{GuoChen2024d}. As such GTEMPO shares a crucial feature with QuAPI and TEMPO: the core algorithms can be read directly from the IF, requiring no further theoretical derivations. 
In fact, most techniques of TEMPO and GTEMPO can be straightforwardly translated into one another. Thanks to this conceptual and algorithmic simplicity, GTEMPO has been successfully applied to solve the AIM on the imaginary-time, Keldysh and the L-shaped Kadanoff contours, generalized to multi-orbital impurities~\cite{ChenGuo2024b,SunGuo2025b}, as well as to study the effects of retarded interaction~\cite{ChenGuo2025} and superconducting baths~\cite{GuoChen2026b}.

\section{The improved TTI IF method}\label{sec:method}

\begin{figure}
  \includegraphics[]{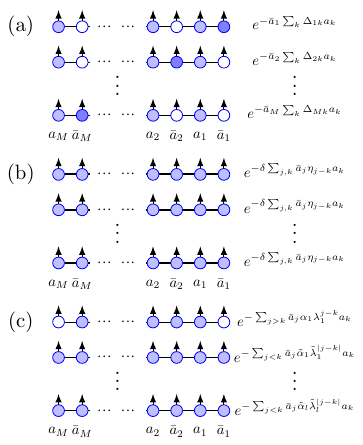} 
  \caption{Schematic illustration of (a) the partial IF method, (b) the TTI IF method and (c) the iTTI IF method. The empty circles mean that these sites (GVs) are not involved in the GMPS, while the light blue circles mean the opposite. The dark blue circles in (a) correspond to the first indices in the partial IF.
    }
    \label{fig:demo}
\end{figure}

For the AIM, the computationally dominant step in all tensor network IF methods is to construct the IF as an MPS. 
In the approach proposed by Abanin \textit{et al.}, a generalized Fishman-White algorithm~\cite{fishman2015-compression,WuStoudenmire2025} is employed to transform the operator $e^{-\sum_{j,k=1}^M \adop_{\uparrow, j}\Delta_{j, k}\adop_{\uparrow, M+k}}$ into a quantum circuit with nearest-neighbor gate operations~\cite{ThoennissAbanin2023b}. The number of gates in the circuit scales roughly as $O[M\log(M)]$, provided that the bipartite entanglement entropy of the underlying Gaussian state grows very slowly, which is claimed to be the case for impurity problems~\cite{ThoennissAbanin2023a}. Consequently, the overall computational cost scales as $O[M\log(M)\chi^3]$. (This estimate essentially counts the number of required singular value decompositions, each with a cost of $O(\chi^3)$).

In GTEMPO, the GT in Eq.(\ref{eq:IF2}) is constructed as a GMPS. To date, two approaches have been developed: (1) the partial IF method~\cite{ChenGuo2024a} and (2) the TTI IF method~\cite{GuoChen2024d}. In the partial IF method, the IF is factorized as
\begin{align}\label{eq:partialif}
e^{-\sum_{j,k=1}^M \abar_{\sigma, j}\Delta_{j, k}a_{\sigma, k}} = \prod_{j=1}^M e^{ -\sum_{k=1}^M \abar_{\sigma, j}\Delta_{j, k}a_{\sigma, k} },
\end{align}
Each term in the product on the right hand side is referred to as a ``partial IF''. Since all terms within a given partial IF share the same GV $\abar_{\sigma, j}$, each partial IF can be exactly represented as a GMPS with bond dimension $2$~\cite{ChenGuo2024b}. The IF can then be constructed by sequentially multiplying these $M$ partial IFs, leading to an overall computational cost of $O(M^2\chi^3)$. Compared to the Fishman-White algorithm, the partial IF method exhibits a less favorable computational scaling with respect to $M$.
However, a crucial advantage of the partial IF method is that the only source of error in the method, on top of the Trotter error from QuAPI, is the MPS bond truncation error, which can be systematically controlled by $\chi$.
The partial IF method is schematically illustrated in Fig.~\ref{fig:demo}(a). We note that both the Fishman-White algorithm and the partial IF method were explored in Ref.~\cite{NgReichman2023}.

In the TTI IF method, one first approximates the hybridization function as a sum of exponential functions 
\begin{equation}\label{eq:expansion}
    \eta_x = \Delta_{i+x, i} \approx
    \left\{
    \begin{array}{ll}
    \sum_{l=1}^n{\alpha_l \lambda_l^x} & \text{if } x > 0; \\[6pt]
    \sum_{l=1}^{n}{\tilde{\alpha}_l \tilde{\lambda}_l^{-x}} & \text{if } x < 0,
    \end{array}
    \right.
\end{equation}
where $\alpha_l$, $\lambda_l$, $\tilde{\alpha}_l$, $\tilde{\lambda}_l$ are complex numbers in general.
Such a decomposition can be obtained, for example, using the Prony method~\cite{marple2019digital}. 
The approximation error occurred in Eq.(\ref{eq:expansion}) is characterized by the mean squared error between the exact and approximated hybridization functions, denoted as $\varsigma$. 
Eq.(\ref{eq:expansion}) also serves as the starting point for HEOM~\cite{YoshitakaKubo1989} and the pseudomode theory~\cite{TamascelliPlenio2018}. Moreover, it is closely connected to the numerical renormalization group method: the logarithmic discretization of the bath typically leads to an exponential decay of the hybridization function~\cite{BullaPruschke2008}, therefore, it could effectively reduce the number of high-frequency pseudomodes (the inverse statement is less clear, i.e, it is not evident whether reducing $n$ in the expansion preferentially eliminates high-frequency pseudomodes). 
From a physical perspective, the long-time behavior is expected to be dominated by the low-frequency pseudomodes associated with smaller $\lambda_l$, as their contributions decay more slowly and thus become increasingly significant at late times.
Furthermore, it has been shown that this expansion converges rapidly under mild continuity conditions on the hybridization function~\cite{ThoennissAbanin2024,VilkoviskiyAbanin2024,HuangLin2026b}, implying that the relevant pseudomodes can be systematically retained with an appropriate implementation.

We rewrite Eq.(\ref{eq:IF2}) as  $\gI_{\sigma}=e^{-\gF_{\sigma}}$, with the exponent $\gF_{\sigma}$ defined as
\begin{align}\label{eq:gF}
\gF_{\sigma} =& \sum_{j,k=1}^M \abar_{\sigma, j} \Delta_{j,k} a_{\sigma, k} \nonumber \\
 =& \sum\limits_{j>k} \bar{a}_{\sigma, j} \eta_{j-k} a_{\sigma, k} + \sum\limits_{j<k} \bar{a}_{\sigma, j} \eta_{j-k} a_{\sigma, k} \nonumber \\
 &+ \sum_{j} \bar{a}_{\sigma, j} \Delta_{j,j} a_{\sigma, j} ,
\end{align}
where the contributions with $j=k$ have been isolated in the third line for later convenience.
Substituting the approximation in Eq.(\ref{eq:expansion}) into the second line of Eq.(\ref{eq:gF}), $\gF_{\sigma}$ can be represented as a GMPS with bond dimension $2n+2$, using the standard technique for representing long-range Hamiltonians as MPOs~\cite{GuoChen2024d}.
Subsequently, the $\WI$ or $\WII$ scheme is employed to approximate $e^{-\delta \gF_{\sigma}}$ (with a step size $\delta$ unrelated to $\beta$ or $t$) as a GMPS with bond dimension $2n+1$~\cite{ZaletelPollmann2015}, and obtain $\gI_{\sigma}$ by multiplying $1/\delta$ such GMPSs. Since all the GMPSs being multiplied are identical, a more efficient scheme can be used which requires only $O[\log(1/\delta)]$ GMPS multiplications~\cite{GuoChen2024d}. The overall computational cost of the TTI IF method thus scales as $O[\log(1/\delta)M\chi^4]$, by using an iterative scheme to variationally fit the result of the multiplication of two GMPSs with bond dimension $\chi$~\cite{Schollwock2011}. 
The TTI IF method is therefore more efficient than the partial IF method for large $M$. However, it introduces two additional sources of error compared to the partial IF method: (1) the approximation error from representing the hybridization function as a sum of exponential functions in Eq.(\ref{eq:expansion}) (which is a well-defined mathematical problem in its own right, and is independent of the subsequent tensor network operations); and (2) the error arising from the $\WI$ or $\WII$ approximation of $e^{-\delta \gF_{\sigma}}$, controlled by $\delta$. The TTI IF method is schematically illustrated in Fig.~\ref{fig:demo}(b).

To this end, we emphasize that the tensor network IF methods are fundamentally different from the pseudomode theory and HEOM, even though all these method can take Eq.(\ref{eq:expansion}) as the starting point and are free from bath discretization error. First, the tensor network IF methods do not have to rely on the approximation in Eq.(\ref{eq:expansion}), for example, the Fishman-White algorithm and the partial IF method avoid this step altogether, in contrast to HEOM and the pseudomode theory. Second, a more fundamental distinction lies in the degrees of freedom involved. The tensor network IF methods yield a temporal MPS that contains only the impurity degrees of freedom, while in the pseudomode theory, one would still solve the many-body (quasi-)Lindblad equation which contains the impurity and the pseudomodes (if one further traces out the pseudomodes in the Lindblad equation, then the method becomes essentially identical to the tensor network IF methods, and the pseudomodes become completely unnecessary). 
Additionally, the tensor network IF methods, especially TEMPO and GTEMPO, are closely related to the recently proposed process tensor framework~\cite{CostaShrapnel2016,PollockModi2018}, as the ADT can be viewed as a representation of the process tensor which encodes the full dynamical information of the impurity~\cite{KeelingReichman2025,GuoChen2026a}.
By comparison, HEOM involves solving a series of coupled equations derived from the IF and auxiliary modes, while the pseudomode theory is formulated in terms of a Lindblad equation for the combined system of impurity and pseudomodes. In this sense, HEOM and the pseudomode theory are more closely related to each other, and their applications are basically limited by the number of pseudomodes~\cite{MullerStrunz2026}.

We now introduce an improved algorithm, termed the improved TTI IF (iTTI IF) method. It is closely related to the TTI IF method, while achieving systematically higher accuracy and efficiency. Substituting Eq.(\ref{eq:expansion}) and Eq.(\ref{eq:gF}) into Eq.(\ref{eq:IF2}) yields
\begin{align}\label{eq:IF3}
    \gI[\boldabar \bolda] \approx &\prod_{l=1}^n e^{-\sum\limits_{j>k} \bar{a}_j \alpha_l \lambda_l^{j-k} a_k} 
    \prod_{l=1}^{n} e^{-\sum\limits_{j<k} \bar{a}_j \tilde{\alpha}_{l} \tilde{\lambda}_{l}^{|j-k|} a_k} \times \nonumber \\ 
    & e^{-\sum_j \bar{a}_j \Delta_{j,j} a_j},
\end{align}
where the spin indices have been omitted for brevity.
The second line can be represented as a GMPS with bond dimension $1$ if $\bar{a}_j$ and $a_j$ are assigned to a single site, or with bond dimension $2$ if they are treated as separate sites. For conceptual clarity, we adopt the former representation, while the latter is used in our practical implementation. Since this local term incurs negligible cost, we focus in the following on the two terms in the first line. For each $l$, the exponent in each term of the two products can be represented as GMPSs with site tensors
\begin{align}\label{eq:ettiif}
    \begin{pmatrix}
    1 & -\alpha \bar{a} & 0 \\
    0 & \lambda & \lambda a \\
    0 & 0 & 1
    \end{pmatrix}
    ;\quad
    \begin{pmatrix}
    1 & \tilde{\alpha} a & 0 \\
    0 & \tilde{\lambda} & \tilde{\lambda} \bar{a} \\
    0 & 0 & 1
    \end{pmatrix},
\end{align}
respectively, where the time indices are omitted due to time-translation invariance, and the subscript $l$ is also omitted for brevity.
Taking the exponential of these two GMPSs (with step size $1$), the corresponding time-evolution operators can be obtained using the $\WII$ approximation~\cite{ZaletelPollmann2015}:
\begin{align}\label{eq:WII}
    \begin{pmatrix}
    1 & -\alpha \bar{a} \\
    \lambda a & \lambda - \alpha \lambda a \bar{a} \\
    \end{pmatrix}
    ;\quad
    \begin{pmatrix}
    1 & \tilde{\alpha} a \\
    \tilde{\lambda} \bar{a} & \tilde{\lambda} - \tilde{\alpha} \tilde{\lambda} a\bar{a} \\
    \end{pmatrix}.
\end{align}
A central result of this work is the following theorem.

\textbf{Theorem 1.} The GMPSs defined by the site tensors in Eq.(\ref{eq:WII}) are the exact exponentials of those in Eq.(\ref{eq:ettiif}). The proof is provided in Appendix.~\ref{app:details}.

We emphasize that this result is highly nontrivial. It is not generalizable to general commuting Hamiltonians with two-body interactions (for example, upon replacing all GVs with Pauli Z operator), 
nor does it seem to be generalizable to the bosonic case (see Appendix.~\ref{app:WII} for details). Similar to the TTI IF method, the iTTI IF method can be directly used to construct an infinite GMPS representation of the IF (infinite along the time axis). For finite $t$, one only needs to impose appropriate boundary conditions on each small GMPS~\cite{ZaletelPollmann2015,GuoChen2024d}. We also note that, although the TTI property appears to be explicitly broken in the partial IF method as can be seen from Eq.(\ref{eq:partialif}), it can be restored by ``rotating'' the 2D tensor network in Fig.~\ref{fig:demo}(a) by $45$ degrees~\cite{LinkStrunz2024}.

Combining Eq.(\ref{eq:IF3}) and Eq.(\ref{eq:WII}), the approximate $\gI$ can be constructed as a product of $2n$ GMPSs for imaginary-time evolution, each with bond dimension $2$. For real-time evolution, there are $8n$ GMPSs to be multiplied, as each GV can reside on either of the two branches~\cite{ChenGuo2024a}. These results further imply that, in the worst case, the bond dimension of the IF scales as $2^{2n}$ and $2^{8n}$ for imaginary- and real-time evolution, respectively. The iTTI IF method is schematically illustrated in Fig.~\ref{fig:demo}(c).
If $n$ scales as $\log(t)$ (as is typically the case for real-time evolution in quantum impurity problems~\cite{VilkoviskiyAbanin2024,ThoennissAbanin2024,HuangLin2026b}), the overall computational cost for the real-time evolution of the AIM grows polynomially with $t$. In practice, we find that the bond dimension can be substantially reduced via MPS bond truncation without compromising accuracy, as demonstrated in the numerical results. The computational cost of the iTTI IF method scales as $O(nM\chi^3)$, arising from the sequential multiplication of a GMPS with bond dimension $\chi$ and $n$ GMPSs with bond dimension $2$. Notably, since $n \propto \log(M)$, this scaling is similar to that of the Fishman-White algorithm, while the iTTI IF method retains explicit time-translationally invariance. 
A brief summarization of various existing methods to build the MPS representation of the IF for the transient real-time dynamics of fermionic impurity problems is shown in Table.~\ref{tab:comparison}.

\begin{table}[!htb]
  \begin{center}
    \caption{The computational costs of various existing algorithms to build the MPS representation of the IF for the transient real-time dynamics of fermionic impurity problems, where $M$ is the total number of discrete time steps, $n$ denotes the number of exponential functions in Eq.(\ref{eq:expansion}).}
    \label{tab:comparison}
    \begin{tabular}{c|c|c}
    \hline
    \hline 
    Algorithm & Computational cost & TTI property \\
    \hline 
    Fishman-White~\cite{ThoennissAbanin2023b} & $O[M\log(M)]\chi^3$ & Not preserved   \\
    \hline 
    Partial IF~\cite{NgReichman2023,ChenGuo2024a} & $O(M^2\chi^3)$ & Not preserved    \\
    \hline 
    TTI IF~\cite{GuoChen2024d} & $O[\log(1/\delta)M\chi^4]$ & Preserved   \\
    \hline 
    iTTI IF (this work) & $O(nM\chi^3)$ & Preserved  \\
\hline
    \end{tabular}
  \end{center}
\end{table}

As our method is inherently TTI, it could also be used to incorporate the TTI property into the infinite MPS representation of the IF, such that one could directly aim at the equilibrium state or steady state without performing the transient dynamics, and the computational cost would no longer be explicitly dependent on $M$.
By exploring the TTI property and the infinite MPS technique, the costs of the TTI IF and iTTI IF methods would scale as $O[\log(1/\delta)\chi^4]$~\cite{GuoChen2024e,GuoChen2024f} and $O(n\chi^3)$ respectively. 
Another approach to explore the TTI property of the IF within the Fishman-White algorithm is by truncating the effective Hamiltonian in the exponent of Eq.(\ref{eq:abaninIF}) with the approximation: $\Delta_{i+x,i}=0$ for $|x|>N_c$, where $N_c$ is a memory size. As a result, the original infinite Hamiltonian with infinite-range interaction becomes another infinite Hamiltonian with range-$N_c$ interaction only, and one could use the infinite time-evolving block decimation (iTEBD) algorithm~\cite{Vidal2007} to build the IF as an infinite MPS with unit cell size $N_c$, whose cost scales as $O[N_c\log(N_c)\chi^3]$ (in comparison, Refs.~\cite{GuoChen2024e,GuoChen2024f} used the infinite density matrix renormalization group algorithm~\cite{McCulloch2008} to directly work with the effective Hamiltonian in Eq.(\ref{eq:gF}) and the unit cell size can be simply chosen to be $1$). 




\section{Numerical results}\label{sec:result}
\begin{figure}[!h]\centering
  \includegraphics{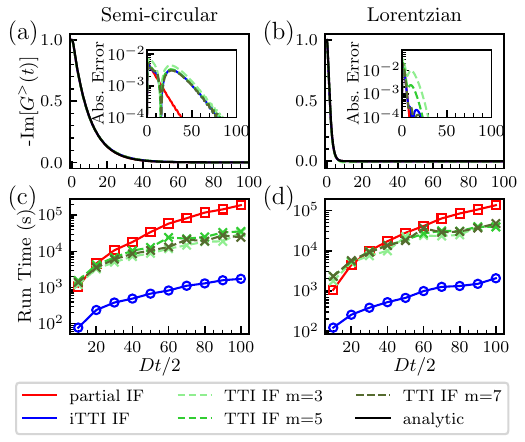}
  \caption{(a,b) Imaginary part of $G^>(t)$ for the semi-circular (a) and Lorentzian (b) BSFs, calculated using the partial IF method (red solid line), the iTTI IF method (blue solid line), and the TTI-IF method with $m=3,5,7$ (green dashed lines from light to dark). The analytical solutions are shown in black solid lines. The insets show the absolute errors of the results calculated using different methods against the analytical solutions. (c,d) The run time of different methods used to construct the IF as GMPSs as a function of the total evolution time $t$ for the semi-circular (c) and Lorentzian (d) BSFs. 
  }
  \label{fig:1}
\end{figure}

\begin{figure}[!h]\centering
  \includegraphics{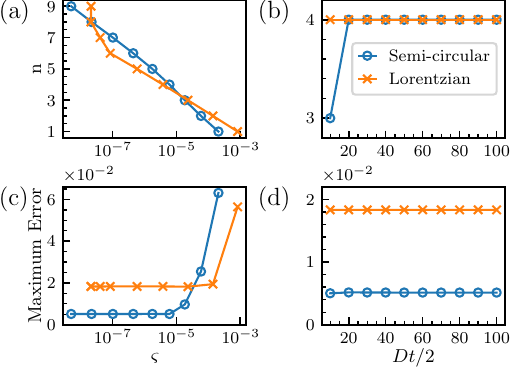}
  \caption{(a) The number of exponential functions $n$ used in Eq.(\ref{eq:expansion}) to achieve a tolerance $\varsigma$ for the semi-circular (blue solid line with circle) and Lorentzian (orange line with x) BSFs, where we have fixed $t=100$. (b) The number of exponential functions $n$ to achieve a fixed tolerance $\varsigma=10^{-5}$ as a function of $t$. (c,d) The maximum error of the imaginary part of $G^>(t)$ calculated using the iTTI IF method against the analytical solutions, as a function of the tolerance $\varsigma$ (c) and $t$ (d), respectively. 
  }
    \label{fig:2}
\end{figure}

\begin{figure}[!h]\centering
  \includegraphics{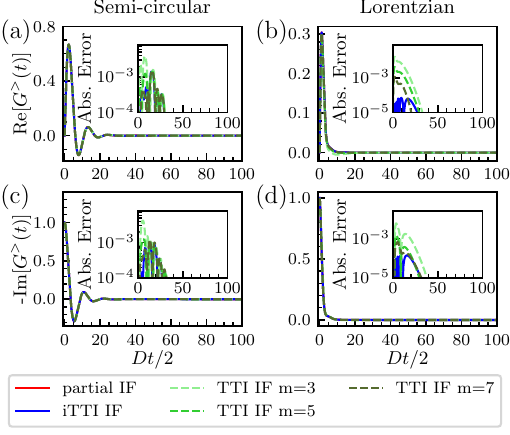}
  \caption{(a,c) Real (a) and imaginary (c) parts of $G^>(t)$ for the AIM with the semi-circular BSF. (b,d) Real (b) and imaginary (d) parts of $G^>(t)$ for the AIM with the Lorentzian BSF. The blue solid lines correspond to the iTTI IF results, while the green dashed lines from light to dark correspond to the TTI IF results with $m=3,5,7$ respectively, similar to Fig.~\ref{fig:1}. The insets show the absolute error of the TTI-IF and iTTI-IF results against the partial IF results.
  }
    \label{fig:3}
\end{figure}

In this section, we demonstrate the efficiency and accuracy of the iTTI-IF method by applying it to the real-time evolution of the AIM.
Although the method can directly access the infinite-time limit, we restrict our numerical simulations to finite times.
We consider an initial state of the form
\begin{align}\label{eq:rho0}
\rhoop_0 = \rhoop_{\rm imp}\otimes \rhoop_{\rm bath}^{\rm th},
\end{align}
where $\rhoop_{\rm imp}$ denotes the impurity density matrix at $t=0$.
To benchmark the overall performance of the iTTI-IF method, we consider two commonly used BSFs: (1) the semi-circular function
\begin{align}\label{eq:semicircular}
    J_s(\epsilon) = \frac{\Gamma}{2\pi}D\sqrt{1-(\epsilon/D)^2},
\end{align}
and (2) the Lorentzian function 
\begin{align}\label{eq:lorentzian}
    J_l(\epsilon) = \frac1\pi \frac{D}{\epsilon^2+D^2}.
\end{align}
In both cases, we set $D=2$ and use it as the unit. 
Unless otherwise specified, in the simulations shown in Fig.~\ref{fig:1}, Fig.~\ref{fig:2} and Fig.~\ref{fig:3}, we use $\Gamma=0.1$, $\beta=10$, $D\delta t=0.1$, and a maximum bond dimension $\chi=80$, and take $\rhoop_{\rm imp}$ to be the vacuum state of the impurity. For this initial condition, the lesser Green’s function $G^<(t) = -\im\langle \adop \aop(t) \rangle$ vanishes. We therefore focus on the greater Green’s function $G^>(t) = -\im\langle \aop(t) \adop \rangle$, which is identical to the retarded Green's function in the present setup. 

To obtain a compact approximation of the hybridization function in Eq.(\ref{eq:expansion}), we employ a scheme which combines the Prony algorithm with a subsequent least-square fitting to minimize $\varsigma$ for a given $n$. Starting from $n=1$, we first apply the Prony algorithm, and then we use its output as the initial guess for the least-square fitting to further reduce $\varsigma$. If the resulting $\varsigma$ exceeds a predefined tolerance, $n$ is increased and the procedure is repeated until the desired accuracy is achieved.

We first apply the iTTI-IF method to the non-interacting case ($U=0$), for which an analytical solution is available. 
We set $\epsilon_d=0$.
Fig.~\ref{fig:1}(a, b) show the imaginary part of $G^>(t)$ as a function of $t$ for the two BSFs defined in Eq.(\ref{eq:semicircular}) and Eq.(\ref{eq:lorentzian}) respectively. The analytical solutions are plotted as black solid lines. Three sets of GTEMPO results are shown, with IF constructed using the partial IF method (red solid line), using the iTTI IF method (the blue solid line), and using the TTI IF method with three different $m=\log_2(1/\delta)$ ($m=3,5,7$ shown as dashed green lines from light to dark). For the approximation in Eq.(\ref{eq:expansion}) which is used in both the TTI-IF and iTTI-IF methods, we impose a tolerance $\varsigma=10^{-5}$, for which $n=4$ is sufficient for both BSFs. The insets of Fig.~\ref{fig:1}(a, b) display the absolute errors of these GTEMPO results relative to the analytical solutions. 
All GTEMPO results show excellent agreement with the analytical solutions for both BSFs, with errors on the order of $10^{-2}$ or below. 
As expected, the partial IF method yields the highest accuracy, as it involves the fewest sources of error.
The TTI IF results approach those of the iTTI IF method as $m$ increases, consistent with the fact that the former incurs an additional time discretization error controlled by $\delta$ (or equivalently $m$).
For sufficiently large $m$, these two sets of results become identical (by design, the TTI IF method converges exponentially fast against $m$~\cite{GuoChen2024d}, which is also reflected in Fig.~\ref{fig:1}(a,b)). 
Fig.~\ref{fig:1}(c,d) show the run time for constructing the IF as GMPSs in these three sets of GTEMPO calculations as a function of the total evolution time $t$. The TTI-IF method becomes more efficient than the partial IF method at large $t$, with only a weak dependence on $m$. Crucially, the iTTI-IF method outperforms both the TTI IF and partial IF methods by at least one order of magnitude, and the computational advantage is expected to become even more pronounced for larger $\chi$ and $M$.

The computational cost and accuracy of the (i)TTI-IF methods are directly related to the number of exponential functions in Eq.(\ref{eq:expansion}), i.e., $n$. In Fig.~\ref{fig:2}(a), we show the value of $n$ required to achieve different tolerance $\varsigma$ at a fixed $t=100$, where the blue solid line with circle and yellow solid line with x represent the results for the semi-circular and Lorentzian BSFs, respectively.
We observe an approximately linear dependence of $n \propto 1/\log(\varsigma)$ for both BSFs (except for the Lorentzian BSF at very small $\varsigma<10^{-7}$). 
In Fig.~\ref{fig:2}(c), we plot the maximum error of $G^>(t)$ obtained using the iTTI IF method, defined as the maximum absolute deviation from the analytical solution over the entire time interval, as a function of the tolerance $\varsigma$ used in Eq.(\ref{eq:expansion}).
The error saturates at $\varsigma = 10^{-5}$ and $\varsigma = 10^{-4}$ for the semi-circular and Lorentzian BSFs, respectively.
In Fig.~\ref{fig:2}(b), we show $n$ as a function of $t$, with a fixed tolerance of $\varsigma = 10^{-5}$ for both BSFs. We observe that $n$ remains nearly constant over the time interval considered. Under the same setup as in Fig.~\ref{fig:2}(b), Fig.~\ref{fig:2}(d) shows the maximum error of $G^>(t)$ calculated by the iTTI-IF method against the analytical solution, as a function of $t$, The maximum error remains on the order of $10^{-2}$ and stays nearly constant throughout the entire time interval.

We next use the IFs constructed above to compute $G^>(t)$ for the half-filled AIM with $U=1$ and $ \epsilon_d=-0.5$. In the absence of an exact analytical solution for the interacting case, the partial IF results are taken as a reference to assess the accuracy of the iTTI IF method and the TTI IF method with $m=3,5,7$. 
Fig.~\ref{fig:3}(a,c) show the real and imaginary parts of $G^>(t)$ for the semi-circular BSF, while Fig.~\ref{fig:3}(b,d) show the corresponding results for the Lorentzian BSF. The iTTI-IF results are shown as blue solid lines, and the TTI IF results with $m=3,5,7$ are shown as green dashed lines from light to dark. The insets show the absolute deviations of these results from the partial IF results. For both BSFs, the iTTI IF results are consistently closest to the partial IF reference, while the TTI IF results systematically approach the iTTI IF results as $m$ increases, similar to the observations in the non-interacting case. 


\begin{figure}[!h]\centering
  \includegraphics{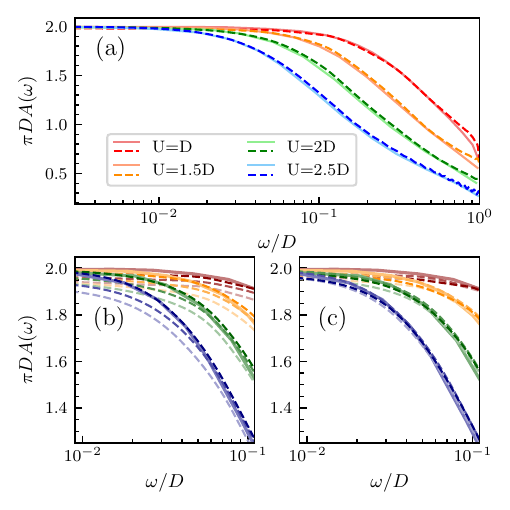}
  \caption{(a) Low-energy spectral function $A(\omega)$ of the AIM under the semi-circular BSF in Eq.(\ref{eq:semicircular}) with $\Gamma=1$ and $\beta=+\infty$. 
  The red, orange, green and blue dashed lines are results calculated using iTTI IF method for $U/D=1,1.5,2,2.5$ respectively, where we have chosen $\chi=240$, 
  $D\delta t=0.02$ and a equilibration time $Dt_0=100$. The solid lines with the same colors are results taken from Ref.~\cite{CaoParcollet2024}.
  (b,c) Zooming in on the spectral function within the frequency range $\omega/D \in [0.01, 0.1]$, where in panel (b) the dashed lines with the same color from lighter to 
  darker are the iTTI IF results for $D\delta t=0.1, 0.04, 0.02$ respectively, with fixed $\chi=240$, while in panel (c) the dashed lines from lighter to darker are 
  the iTTI IF results for $\chi=200, 240, 280$ respectively, with fixed $D\delta t=0.02$.
  }
    \label{fig:4}
\end{figure}

As an application of the iTTI-IF method, we calculate the spectral function for the model considered in Ref.~\cite{CaoParcollet2024}, i.e., the half-filling AIM ($\epsilon_d=-U/2$) under a semi-circular BSF with $\Gamma=1$ and $\beta=+\infty$ (zero temperature). 
The spectral function is defined as 
\begin{align}
\pi A(\omega) = -{\rm Im}[G^R(\omega)],
\end{align}
where $G^R(\omega)$ is the Fourier transform of the retarded Green's function $G^R(t) = G^>(t) - G^<(t)$. The same problem is also consider in Ref.~\cite{SonnerAbanin2025} for a fixed $U=2D$. In our simulations of this problem, we choose $\rhoop_{\rm imp}$ to be the local thermal state, i.e., $\rhoop_{\rm imp} = e^{-\beta\Himp}$, to accelerate convergence to equilibrium, we set an equilibration time $Dt_0 = 100$ (which means that we assume the impurity and bath reach equilibrium after $t_0$) and calculate the equilibrium Green's functions thereafter~\cite{ChenGuo2024c}.
We note that the results in Ref.~\cite{CaoParcollet2024} are obtained using a complex-time evolution scheme, while our approach is based on pure real-time evolution.
The spectral functions for $U/D=1,1.5,2,2.5$ are shown in Fig.~\ref{fig:4}(a). The dashed lines with different colors 
correspond to our results for different values of $U$, calculated using $n=6$, $\chi=240$ and $D\delta t=0.02$, while the 
solid lines with the same colors are results from Ref.~\cite{CaoParcollet2024}.
Overall, the two sets of results are in good agreement. There are some deviations at large $\omega$, which are likely 
due to the finite values of $\delta t$ and $t_0$. 
In Fig.~\ref{fig:4}(b,c), we show the convergence of our iTTI IF results against the two hyperparameter $\delta t$ and $\chi$ respectively.
From Fig.~\ref{fig:4}(b), we can see that the choice of $\delta t$ significantly affects the accuracy of iTTI IF. Our results become closer 
to the reference data for smaller $\delta t$ and match well with it at $D\delta t=0.02$. 
From Fig.~\ref{fig:4}(c), we can see that our results improve when increasing $\chi$ from $200$ to $240$,
and then almost saturate when further increasing $\chi$.
In these calculations, $G^R(t)$ is evaluated up to $D(t-t_0)=100$, corresponding to a total evolution time of $Dt=200$. 
Moreover, the small time step $D\delta t=0.02$ is required to ensure convergence for all different values of $U$, making the simulation computationally quite 
expensive (obtaining the IF as a GMPS takes $80$ hours using a single thread, which is almost impossible using the partial IF method due to the large value of $M$). 
For even larger values of $U$ as considered in Ref.~\cite{CaoParcollet2024}, the present setup does not yield converged results with the chosen $t_0$, $\chi$ and $\delta t$ (for $U=2.5D$ we already see strange small oscillations at large frequency). 
This is expected, as larger $U$ typically requires larger $t_0$ to reach equilibrium and a smaller $\delta t$ for numerical stability. 
This highlights the intrinsic difficulty of the problem, particularly at large $U$, which is also the reason that the complex-time evolution scheme is used in Ref.~\cite{CaoParcollet2024}. 
Incorporating complex-time evolution schemes into the GTEMPO framework is straightforward and could be an interesting direction for future investigation.

\section{Conclusion}{\label{sec:summary}}
In summary, we have proposed an improved TTI IF method to efficiently and accurately construct the Feynman-Vernon IF as a Grassmann MPS, which is the computationally dominant step in the GTEMPO approach to solve the AIM. 
The iTTI IF method achieves a lower computational cost compared to the existing alternatives to build the MPS representation of the IF whiling maintaining the TTI property. 
Compared to the partial IF method, the only additional source of error in the iTTI IF method arises from approximating the hybridization function as a sum of exponential functions. Compared to the TTI IF method, the iTTI IF method fully exploits the structure of the Grassmann algebra, and is both more efficient and more accurate, as it involves fewer sources of error. Similar to the TTI IF method, the iTTI IF method can be straightforwardly applied to both finite- and infinite-time evolution. In our numerical benchmarks, we consider two commonly used bath spectral functions, and both the non-interacting and interacting cases. The results demonstrate that the iTTI IF method achieves an accuracy comparable to that of the partial IF method, while consistently outperforming the TTI IF method. Crucially, we find that the iTTI IF method is at least one order of magnitude faster than the partial IF and the TTI IF methods. 
These results establish the iTTI-IF method as an efficient tool for accelerating simulations of the AIM.

For multi-orbital extensions of the AIM, the iTTI IF method can still be used to accelerate the construction of the GMPS representation of the IF of each flavor. However, in this case the overall computational cost may instead be dominated by the evaluation of the ADT or observables~\cite{ChenGuo2024b}.

\begin{acknowledgments}
We thank Yi Lu and Xiaodong Cao for the insightful discussions, and for providing us the data in Fig.~\ref{fig:4} for benchmarking.
Z. L. is partially supported by NSFC (22393913), by the Strategic Priority Research Program of the Chinese Academy of Sciences (XDB0450101). 
\end{acknowledgments}

\appendix

\section{Overview of the $\WI$ and $\WII$ approximations}
\label{app:WII}

\setlength{\tabcolsep}{4pt}
\begin{table*}[!t]
\centering
\caption{Three commuting Hamiltonians with exponentially decaying interactions, together with their corresponding $\WI$ and $\WII$ approximations. $\hat{H}_1$ represents a long-range Ising chain. $\hat{H}_3$ is obtained by mapping the Hamiltonian in Eq.(\ref{eq:ettiif}) to an equivalent spin model via the Jordan-Wigner transformation, where the second subscript is because the GVs appear in conjugate pairs. $\hat{H}_2$ is a simplified version of $\hat{H}_3$ if the Hamiltonian in Eq.(\ref{eq:ettiif}) only contains $a$ or $\abar$. The first index of the operators in the last three columns are omitted due to the translational invariance.}
\label{tab:1}
\begin{tabular}{cccc}
\toprule
Hamiltonian & Site tensor & $\WI$ & $\WII$ \\
\midrule

$\hat{H}_1 = \sum_{i<j} \lambda^{|i-j|-1} \hat{Z}_i \hat{Z}_j$ & 
$\begin{pmatrix}
\hat{I} & \hat{Z} & 0 \\
0 & \lambda \hat{I} & \hat{Z} \\
0 & 0 & \hat{I}
\end{pmatrix}$ &
$\begin{pmatrix}
\hat{I} & \sqrt{\tau}\hat{Z} \\
\sqrt{\tau}\hat{Z} & \lambda \hat{I} 
\end{pmatrix}$ &
$\begin{pmatrix}
\hat{I} & \sqrt{\tau}\hat{Z} \\
\sqrt{\tau}\hat{Z} & (\lambda+\tau) \hat{I} 
\end{pmatrix}$ \\

$\hat{H}_2 = \sum_{i<j} \lambda^{|i-j|-1} \hat{Q}^\dagger_i \prod_{i<k<j} \hat{Z}_k \hat{Q}^\dagger_j$ &
$\begin{pmatrix}
\hat{I} & \hat{Q}^\dagger & 0 \\
0 & \lambda \hat{Z} & \hat{Q}^\dagger \\
0 & 0 & \hat{I}
\end{pmatrix}$ &
$\begin{pmatrix}
\hat{I} & \sqrt{\tau}\hat{Q}^\dagger \\
\sqrt{\tau}\hat{Q}^\dagger & \lambda\hat{Z} 
\end{pmatrix}$ &
$\begin{pmatrix}
\hat{I} & \sqrt{\tau}\hat{Q}^\dagger \\
\sqrt{\tau}\hat{Q}^\dagger & \lambda\hat{Z} 
\end{pmatrix}$ \\

$\begin{aligned} \hat{H}_3 = \sum_{i<j}& \lambda^{|i-j|-1} \hat{Q}^\dagger_{i,1}\hat{Z}_{i,2} \cdot \\ & \prod_{i<k<j}(\hat{Z}_{k,1}\hat{Z}_{k,2}) \hat{Z}_{j,1}\hat{Q}^\dagger_{j,2} \end{aligned}$ &
$\begin{pmatrix}
\hat{I} & \hat{Q}^\dagger_{,1} \hat{Z}_{,2} & 0 \\
0 & \lambda \hat{Z}_{,1} \hat{Z}_{,2} & \hat{Z}_{,1} \hat{Q}^\dagger_{,2} \\
0 & 0 & \hat{I}
\end{pmatrix}$ &
$\begin{pmatrix}
\hat{I} & \sqrt{\tau} \hat{Q}^\dagger_{,1} \hat{Z}_{,2} \\
\sqrt{\tau} \hat{Z}_{,1} \hat{Q}^\dagger_{,2} & \lambda \hat{Z}_{,1} \hat{Z}_{,2} 
\end{pmatrix}$ &
$\begin{pmatrix}
\hat{I} & \sqrt{\tau} \hat{Q}^\dagger_{,1} \hat{Z}_{,2} \\
\sqrt{\tau} \hat{Z}_{,1} \hat{Q}^\dagger_{,2} & \lambda \hat{Z}_{,1} \hat{Z}_{,2} - \tau \hat{Q}^\dagger_{,1} \hat{Q}^\dagger_{,2}
\end{pmatrix}$ \\

\bottomrule
\end{tabular}
\end{table*}


In this section we briefly review the $\WI$ and $\WII$ schemes introduced in Ref.~\cite{ZaletelPollmann2015}, which provide first-order approximations to the exponential of a Hamiltonian represented as an MPO. 
We consider a translationally-invariant Hamiltonian $\hat{H}$ that is represented as an MPO with site tensor
\begin{align}
    \begin{pmatrix}
    \hat{I} & \bm{\hat{C}}^T & \hat{D} \\
    0 & \bm{\hat{A}} & \bm{\hat{B}} \\
    0 & 0 & \hat{I}
    \end{pmatrix},
\end{align}
where $\hat{I}$ denotes the identity operator, $\hat{D}$ corresponds to the local term in $\hat{H}$, $\bm{\hat{B}} $ and $\bm{\hat{C}} $ are operator-valued vectors encoding inter-site couplings ($\bm{x}^T$ means the transpose of the vector $\bm{x}$), $\bm{\hat{A}}$ is an operator-valued matrix that encodes long-range interactions.

The $\WI$ approximation to $e^{\hat{H}\tau}$ is given by
\begin{align}
    \begin{pmatrix}
    \hat{I} + \tau\hat{D} & \sqrt{\tau}\bm{\hat{C}}^{T} \\
    \sqrt{\tau}\bm{\hat{B}} & \bm{\hat{A}} 
    \end{pmatrix},
\end{align}
which retains only the non-overlapping terms in the Taylor expansion of $e^{\hat{H}\tau}$, such as $\bm{\hat{C}}_1^T \bm{\hat{A}}_2 \bm{\hat{B}}_3$, $\bm{\hat{C}}_1^T \bm{\hat{A}}_2 \bm{\hat{B}}_3 \cdot \bm{\hat{C}}^T_6 \bm{\hat{B}}_7$, $\bm{\hat{C}}^T_1 \bm{\hat{A}}_2 \bm{\hat{B}}_3 \cdot \hat{D}_4 \cdot \bm{\hat{C}}_6^T \bm{\hat{B}}_7$.

The $\WII$ scheme provides a more accurate first-order approximation than $\WI$. For commuting Hamiltonians, the $\WII$ approximation contains all non-bond overlapping terms, i.e., any two factors in the product overlap on no more than one site, for example $\bm{\hat{C}}^T_1 \bm{\hat{A}}_2 \bm{\hat{B}}_3 \cdot \bm{\hat{C}}_3^T \bm{\hat{B}}_4$, $\bm{\hat{C}}_1^T \bm{\hat{A}}_2 \bm{\hat{B}}_3 \cdot \hat{D}_2 \cdot \bm{\hat{C}}^T_3 \bm{\hat{B}}_4$. However, terms which overlap on more than one site are not included, such as $\bm{\hat{C}}_1^T \bm{\hat{A}}_2 \bm{\hat{B}}_3 \cdot \bm{\hat{C}}_2^T \bm{\hat{B}}_3$. 
The site tensor in the $\WII$ approximation can be written as
\begin{align}
    \begin{pmatrix}
    \hat{W}_D & \bm{\hat{W}}_C^T \\
    \bm{\hat{W}}_B & \bm{\hat{W}}_A
    \end{pmatrix},
\end{align}
where the sub-blocks are defined by the following Taylor expansion:
\begin{align}
    &e^{\bm{\phi}^T \cdot \bm{\hat{A}} \cdot \bar{\bm{\phi}} + 
    \bm{\phi}^T \cdot \bm{\hat{B}} \sqrt{\tau} +
    \sqrt{\tau} \bm{\hat{C}}^T \cdot \bar{\bm{\phi}} +
    \tau\hat{D}} \nonumber \\
    =& \hat{W}_D +
    \bm{\hat{W}}_C^T \cdot \bar{\bm{\phi}} +
    \bm{\phi}^T \cdot \bm{\hat{W}}_B + 
    \bm{\phi}^T \cdot \bm{\hat{W}}_A \cdot \bar{\bm{\phi}} +
    \dots.
\end{align}
Here $\bm{\phi}$ and $\bar{\bm{\phi}}$ are two vector fields.

In Table~\ref{tab:1}, we present the explicit forms of the $\WI$ and $\WII$ approximations for three commuting Hamiltonians with exponentially decaying interactions. Here we have used
\begin{align}
    \hat{Z} = \begin{pmatrix} 1 & 0 \\ 0 & -1 \end{pmatrix},\ \ 
    \hat{Q}^\dagger = \frac12(\hat{X}-\im\hat{Y}) = \begin{pmatrix} 0 & 0 \\ 1 & 0 \end{pmatrix},
\end{align}
where $\hat{X}$, $\hat{Y}$ and $\hat{Z}$ are the Pauli operators.
It is straightforward to verify that the $\WII$ approximation is not exact for the long-range Ising Hamiltonian $\hat{H}_1$ (which illustrates that the Theorem in the main text does not hold for general commuting Hamiltonians). For example, consider a two-site case with $L=2$ for the Hamiltonian $\hat{H}_1$, which simplifies to $\hat{H}_1=\hat{Z}_1\hat{Z}_2$. The $\WII$ approximation of $e^{\tau \hat{H}_1}$ is $\hat{I}+\tau \hat{Z}_1\hat{Z}_2$, whereas the exact exponential is $\cosh(\tau)\hat{I} + \sinh(\tau)\hat{Z}_1\hat{Z}_2$. 
In contrast, the $\WII$ approximations for $\Hop_2$ and $\Hop_3$ are exact. These Hamiltonians can be viewed as analogies of the Grassmann tensor in Eq.(\ref{eq:gF}), obtained via the Jordan-Wigner transformation.
Moreover, we can see that the $\WI$ and $\WII$ approximations for $\hat{H}_2$ are equal to each other, since the one-site overlapping terms vanish due to the relation ${ (\hat{Q}^\dagger )}^2 = 0$. 


\section{Proof of Theorem 1}\label{app:details}

Since the two GMPSs in Eq.(\ref{eq:ettiif}) have essentially the same form, it suffices to prove Theorem 1 for one of them. 
In the following, we focus on the first product in the first line of Eq.(\ref{eq:IF3}).
We begin by performing a Taylor expansion of a single term in this product, 
\begin{align}\label{eq:taylor}
    e^{-\sum\limits_{j>k} \bar{a}_j \alpha \lambda^{j-k} a_k}
    =& 1 - {\sum\limits_{j>k} \bar{a}_j \alpha \lambda^{j-k} a_k} \nonumber \\
    &+ \frac{1}{2!} \left( {\sum\limits_{j>k} \bar{a}_j \alpha \lambda^{j-k} a_k} \right)^2 - \cdots ,
\end{align}
where the subscript $l$ has been omitted for brevity.
It has been shown that the $\WII$ approximation contains all non-bond overlapping terms in the Taylor expansion for commuting Hamiltonians~\cite{ZaletelPollmann2015}.
Therefore the zeroth- and first-order terms on the first line of Eq.(\ref{eq:taylor}) are already captured by the $\WII$ approximation. In the following, we show that all bond overlapping terms in the Taylor expansion cancel each other exactly, thereby completing the proof.

We adopt a normal-ordering convention consistent with that used in our numerical implementation, in which the GVs are ordered according to their time step indices. For GVs sharing the same index, $a$ is placed before $\bar{a}$ by convention. For each resulting pattern (i.e. a normal ordered sequence of GVs), multiple combinations of the quadratic terms in the exponent may contribute. To account for all such combinations, one must systematically enumerate all possible pairings of $\bar{a}_i a_j$ with $i>j$ that generate the given pattern. For a given pattern, different pairings yield the same scalar prefactor; only the anti-commutation relations between the GVs can introduce different sign factors. Therefore, we omit the common scalar prefactor in the following for simplicity.
If two $a$ (or $\bar{a}$) share the same index , this pattern vanishes due to the Grassmann algebra $a^2=\bar{a}^2=0$. So we will omit these trivially vanishing cases. We note that with our convention, a legal pattern must start with $\abar$ and end with $a$.

As illustrative examples, we analyze the second- and third-order contributions. For each order, we enumerate all possible normal-ordered patterns and determine whether they yield non-vanishing contributions.

At second order, there are only two possible patterns, i.e., $\bar{a}_i a_j \bar{a}_k a_l$ and $\bar{a}_i \bar{a}_j a_k a_l$. We categorize them into two sets:
\begin{itemize}
    \item Set-1 $S_1 = \{\bar{a}_i a_j \bar{a}_k a_l\}$. The only possible pairing for this pattern is of the form $\bar{a}_i a_j * \bar{a}_k a_l$ (denoted as $\wick{\c1 {\bar{a}_i} \c1 a_j \c1 {\bar{a}_k} \c1 a_l }$), which is already contained in the $\WII$ approximation, as the two pairings overlap on at most one site (specifically when $j=k$).
    
    \item Set-2 $S_2 = \{\bar{a}_i \bar{a}_j a_k a_l\}$, which can arise from $\bar{a}_i a_k * \bar{a}_j a_l$ ($\wick{\c2 {\bar{a}_i} \c1 {\bar{a}_j} \c2 a_k \c1 a_l}$) or $\bar{a}_i a_l * \bar{a}_j a_k$ ($\wick{\c2 {\bar{a}_i} \c1 {\bar{a}_j} \c1 a_k \c2 a_l}$). These two combinations carry opposite sign factors (the former requires one swap operation, while the latter requires two), and therefore cancel each other exactly.
\end{itemize}

Similarly, at third order, we classify all possible patterns into two sets:
\begin{itemize}
    \item Set-1 $S_1 = \{\bar{a}_i a_j \bar{a}_k a_l \bar{a}_m a_n\}$. As in the second-order case, the only possible pairing is of the form $\bar{a}_i a_j * \bar{a}_k a_l * \bar{a}_m a_n$ ($\wick{\c1 {\bar{a}_i} \c1 a_j \c1 {\bar{a}_k} \c1 a_l \c1 {\bar{a}_m} \c1 a_n}$), which is already contained in the $\WII$ approximation.
    \item Set-2 $S_2$ contains all remaining patterns, which can be exhaustively listed as
    \begin{align}
    S_2 = \{&\bar{a}_i a_j \bar{a}_k \bar{a}_l a_m a_n, \bar{a}_i \bar{a}_j a_k \bar{a}_l a_m a_n, \nonumber \\ 
    &\bar{a}_i \bar{a}_j \bar{a}_k a_l a_m a_n, \bar{a}_i \bar{a}_j a_k a_l \bar{a}_m a_n\}.
    \end{align}
     We analyze each pattern in $S_2$ as follows.
    \begin{itemize}
        \item $\bar{a}_i a_j \bar{a}_k \bar{a}_l a_m a_n$. There are many possible pairings to generate this pattern. In order to enumerate all of them, we first group the last $a$ with one preceding $\bar{a}$, and if left terms form a legal pattern, this pairing is legal and will be considered. This yields two pairings: $\bar{a}_i a_j \bar{a}_l a_m * \bar{a}_k a_n$ ($\wick{\c1 {\bar{a}_i} \c1 a_j \c2 {\bar{a}_k} \c1 {\bar{a}_l} \c1 a_m \c2 a_n}$) and $\bar{a}_i a_j \bar{a}_k a_m * \bar{a}_l a_n$ ($\wick{\c1 {\bar{a}_i} \c1 a_j \c1 {\bar{a}_k} \c2 {\bar{a}_l} \c1 a_m \c2 a_n}$), which cancel with each other. Here we note that $a_j \bar{a}_k \bar{a}_l a_m * \bar{a}_i a_n$ is an illegal pairing.
        
        \item $\bar{a}_i \bar{a}_j a_k \bar{a}_l a_m a_n$. There are three possible pairings: $\bar{a}_j a_k \bar{a}_l a_m * \bar{a}_i a_n$ ($\wick{\c2 {\bar{a}_i} \c1 {\bar{a}_j} \c1 a_k \c1 {\bar{a}_l} \c1 a_m \c2 a_n}$), $\bar{a}_i a_k \bar{a}_l a_m * \bar{a}_j a_n$ ($\wick{\c1 {\bar{a}_i} \c2 {\bar{a}_j} \c1 a_k \c1 {\bar{a}_l} \c1 a_m \c2 a_n}$) and $\bar{a}_i \bar{a}_j a_k a_m * \bar{a}_l a_n$ ($\wick{\bar{a}_i \bar{a}_j a_k \c1 {\bar{a}_l} a_m \c1 a_n}$). The first two cancel each other. The third contains the sub-pattern $\bar{a}_i \bar{a}_j a_k a_m$, which belongs to $S_2$ at second order, so it has zero contribution upon further decomposing into $\bar{a}_j a_k * \bar{a}_i a_m * \bar{a}_l a_n$ ($\wick{\c2 {\bar{a}_i} \c1 {\bar{a}_j} \c1 a_k \c1 {\bar{a}_l} \c2 a_m \c1 a_n}$) and $\bar{a}_i a_k * \bar{a}_j a_m * \bar{a}_l a_n$ ($\wick{\c1 {\bar{a}_i} \c2 {\bar{a}_j} \c1 a_k \c1 {\bar{a}_l} \c2 a_m \c1 a_n}$).
        
        \item $\bar{a}_i \bar{a}_j \bar{a}_k a_l a_m a_n$. There are three possible pairings: $\bar{a}_j \bar{a}_k a_l a_m * \bar{a}_i a_n$ ($\wick{\c1 {\bar{a}_i} \bar{a}_j \bar{a}_k a_l a_m \c1 a_n}$), $\bar{a}_i \bar{a}_k a_l a_m * \bar{a}_j a_n$ ($\wick{\bar{a}_i \c1 {\bar{a}_j} \bar{a}_k a_l a_m \c1 a_n}$) and $\bar{a}_i \bar{a}_j a_l a_m * \bar{a}_k a_n$ ($\wick{\bar{a}_i \bar{a}_j \c1 {\bar{a}_k} a_l a_m \c1 a_n}$). They all contain the $\bar{a} \bar{a} a a$ sub-pattern, that belongs to the $S_2$ at second order, thus they all have zero contributions upon further decomposition.
        
        \item $\bar{a}_i \bar{a}_j a_k a_l \bar{a}_m a_n$. The only possible pairing in this pattern is of the form $\bar{a}_i \bar{a}_j a_k a_l * \bar{a}_m a_n$ ($\wick{\bar{a}_i \bar{a}_j a_k a_l \c1 {\bar{a}_m} \c1 a_n}$), which also have zero contribution as it contains the $\bar{a} \bar{a} a a$ sub-pattern.
    \end{itemize}
\end{itemize}

We therefore classify all the patterns at each order into two sets:
\begin{itemize}
    \item Set-1 ($S_1$), which contains a single pattern with a strictly alternating ordering of the form $\bar{a} a \bar{a} a \dots \bar{a} a$, which is already included in the $\WII$ approximation.  
    \item Set-2 ($S_2$), which contains all remaining patterns, characterized by the presence of multiple overlapping sites and are not captured by the $\WII$ approximation.  
\end{itemize}

\textbf{Lemma.} At any order $n$, every pattern that belongs to Set-2 has zero contribution.

Proof of Lemma. The statement holds for $n=2$ as shown above. Assume that it holds at order $n$. We then consider a pattern in Set-2 at order $n+1$. Grouping the last $a$ with one preceding $\bar{a}$, if the remaining GVs form a Set-2 pattern at order $n$, the result vanishes by the induction hypothesis, if the remaining part is in Set-1 instead, then the pattern must necessarily be of the form
\begin{align}
    (\bar{a} a)^n \bar{a} \bar{a} a (\bar{a} a)^m a, \ \ \ n,m\in\{0,1,2\dots\}.
\end{align}
There are always two combinations that result in the sub-pattern $\bar{a} a \bar{a} a \dots \bar{a} a$:
\begin{align}
    \wick{(\bar{a} a)^n \c1 {\bar{a}} \bar{a} a (\bar{a} a)^m \c1 a}, \ \ \ \wick{(\bar{a} a)^n \bar{a} \c1 {\bar{a}} a (\bar{a} a)^m \c1 a},
\end{align}
which cancel each other exactly. Thus all Set-2 contributions vanish at order $n+1$, completing the induction.

Since all patterns in Set-2 vanish, the $\WII$ approximation is exact for the GMPSs in Eq.(\ref{eq:ettiif}), thereby completing the proof of Theorem~1. Finally, we note that this proof does not seem to be generalizable to the bosonic case. As is evident from the proof, the anti-commutation relations of the GVs play a crucial role, and such a structure is absent in the bosonic case.

\bibliography{refs}

\end{document}